# Using math in physics:
# 4. *Toy models*

*Edward F. Redish,*
University of Maryland - emeritus, College Park, MD

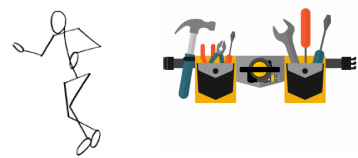

Learning to create, use, and evaluate models is a central element of becoming a scientist. In physics, we often begin an analysis of a complex system with highly simplified or *toy models*. In introductory physics classes, we tend to use them without comment or motivation. Some students infer that physics is irrelevant to their understanding of the real world and are discouraged from making the cognitive blend of physics concepts with math symbology essential for making sense of physics. In this paper, I discuss the often hidden barriers that make it difficult for our students to accept and understand the value of toy models, and suggest instructional approaches that can help.

Simplified toy models fill our introductory physics classes. We discuss the motion of a point projectile in a vacuum using flat-earth gravity, the rotation of perfectly rigid bodies, infinite parallel plate capacitors, resistanceless wires, and many others. Toy models occur at the graduate and even the professional level, where we consider perfectly symmetric infinite situations to teach Gauss's and Ampere's law, the scattering of a quantum electron by a delta-function potential in 1D, the atomic shell model, the $\Phi^4$ relativistic quantum field theory, the Ising model, and the Schwartzschild solution to Einstein's general relativity equations.

As physicists, we consider our highly simplified models an obvious and natural way to approach physics. Mathematical models of complex systems can be tricky, so the best way to understand the math is to take the simplest possible example that illustrates a phenomenon, then take it apart and put it back together again, matching the math with physical intuitions and building a mental blend of what the math means physically. Knowing the value of this (and other knowledge-building tools) is *an epistemological resource*: (0) *Simple systems help build understanding.* Learning to use this resource effectively to build new understanding is an important step in learning to be an effective scientist.

This paper is part of a series of papers on "Using math in physics."[1] Each focuses on a particular approach or strategy ("epistemic game") that can be a part of helping students to learn to blend physical concepts, knowledge, and intuition with mathematical symbols and processing.

Toy models are one such strategy: highly simplified physical situations that ignore many factors that are present in real situations in order to gain insight into how the math represents the physics, one factor at a time. The icon I use for the toy model game is the stick figure shown at the top, representing a bare outline without details. I use this in class to provide a visual marker to remind students how valuable (and common) this strategy is.

There is increasing awareness of the importance of discussing modeling in introductory physics classes at the high school level, thanks largely to the work of David Hestenes, Jane Jackson, the modeling group,[2] and their widespread instructional model for high school teachers. But if you're not teaching a modeling class, how explicit do you need to be about the role of modeling?

## We need to be explicit about our use of toy models in introductory physics

An understanding of the value and utility of toy models is not an epistemological resource that many of our students bring to an introductory level physics class. It can be particularly difficult for non-majors.[3]

When our students come from another discipline, they may bring disciplinary epistemological resources that can cause them to be uncomfortable with the ones we are hoping they will learn. For example, my life-science students bring two that suggest to them that toy models are misleading: (1) *Life is complex* — Living organisms require multiple related and coordinated processes to maintain life; and (2) *Structure determines function* — The historical fact of natural selection leads to strong relationships between the structure of living organisms and how they function. If you try to simplify an organism, it dies.

Many biology students activate these resources in physics when confronted with our highly simplified toy models and, as a result, are dismissive of them, not seeing the value that we take for granted. (And they rarely recognize as toy models the highly simplified models that are used in introductory biology, such as Punnett squares, phylogenetic trees, or lock-and-key models of enzymes.)

Although a resistance to toy models can be especially strong with life science majors who tend not to use a lot of math in their introductory classes, engineering majors can also show a negative reaction. They tend to be motivated by practical applications and may bring the epistemological resource: (3) *Value is determined by how useful it is.* They see our toy models as overly simplified and theoretical. Physicists are often mocked by non-physicists for working with "spherical cows"[4] or "in frictionless vacuums."[5]

To help our students understand the value of toy models, we need to explain and motivate them, not simply treat them as if they were the natural and obvious way to mathematize the world.

## The structure of mathematical toy models

When we make a mathematical model of a physical system, we blend two mental spaces: our conceptual knowledge of the physical world and the symbolic structures used in math. Fig. 1 shows one way to begin the blend of a physical system with a mathematical representation.[6]

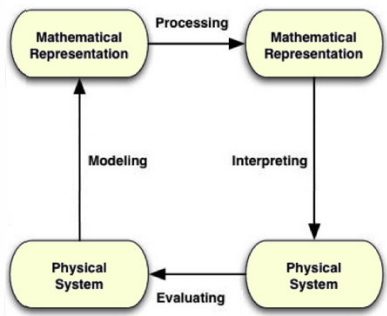

*Figure 1. A simple model of mathematical modeling.*

We can begin in the lower left corner, recognizing some properties of a physical quantity that look like math. Something physical can be quantified (assigned a number or numbers) through a measurement or combination of measurements.[7] For example, we might decide a physical concept can be represented by positive or negative numbers, like a temperature, or that it has both a magnitude and direction and so should be represented by three numbers —a vector, like a velocity. This identification of the physical quantities with mathematical structures is a *mathematical model*.

In choosing a model, we have to decide what phenomena we are trying to describe, how to quantify the quantities involved, and, perhaps most important, what matters and what doesn't. The world is too complex for us to include everything that's going on. Deciding what matters and what can be ignored (at least at first) is an essential scientific skill, one that is, unfortunately, rarely taught explicitly even to our physics majors. It involves a number of our e-games, particularly *estimation*[8] and the development of associated intuitions of scale.

Once we've mapped our physical quantities onto math, we inherit *processing* tools from mathematics that let us solve problems that we might have difficulty solving. But once we have completed our calculation, we have to *interpret* the result back in the physics. What did the solution tell us about the physical world? Finally, we have to *evaluate* that interpretation. Is our model good enough for what we needed to do? Or are there refinements that we have to make, additional factors or effects that we really need to include?

Most of the problems that you'll find in the back of an introductory physics textbook live on the top leg of my modeling diagram: processing only. Students can be misled to think that physics is just math, The other legs are where students have to focus on "the physics" and learn to develop modeling skills and build the blend.

The best way to convince our students of the value of toy models is to be explicit about our modeling process throughout, to explain that toy models aren't the final answer, but only the start of a continuing process of learning to ask the right questions.

## Why are toy models useful and important?

We use toy models widely in introductory physics because they support multiple pedagogically valuable developments.

- Toy models help students build the blend by focusing on the math-physics connection
- Toy models are built into most of our problems and can help build physical intuition
- Some toy models work *waaay* better than we might expect.

*Toy models help students build the blend by focusing on the math-physics connection*

There are many places in introductory physics where what seem to be ridiculously over-simplified toy models can help students understand what's going on and see how the math represents the physics. One example is the constant acceleration, linearly changing velocity example with race cars given in the Anchor Equations paper.[9] There, students have to use the basic kinematics equations and map a physical situation into them for two different objects, helping to build the blend. A more realistic model with non-uniform acceleration would not be appropriate for introductory students to deal with.[10]

Another bald-faced ridiculous example that turns out to be of great value is the infinite sheet of perfectly smooth and uniform charge. This is of great value in the context of the parallel plate capacitor, where it essentially turns complex $1/r^2$ problems into the analog of flat-earth gravity: a constant field. This example permits students to learn how energy can be stored in the separation of charge, to model the forces that keep current going through a resistor, and even to estimate the electrical energy in nerve cells. A way to teach this example as a model is discussed in more detail below.

*Toy models are built into most of our problems and can help build physical intuition*

My grammar checker wants me to use "physics" instead of "the physics" in the last section, but as a professional physics researcher, I often hear someone ask, "What's the physics?" or "We need to get the physics right." In physics lingo, *the physics* means those factors that are essential to understanding the phenomenon at hand. This is where the skill and art of the professional scientist lies: deciding what matters and what can be ignored at first. Building this skill requires the development of a good physical intuition — a sense for quantification and scale that can be helped by learning to do estimations.[11] Many of the standard problems we give in introductory physics have implicit (and unmotivated) toy models. An example is given in the section on using toy models in instruction.

*Some toy models work waaaay better than we expect*

If I hold a 10 cm cast iron sphere above my head to a height of about 3 meters and drop it, it will hit the ground in about 0.8 seconds. If I hold a Styrofoam ball of the same size and drop them together, the Styrofoam takes noticeably longer to fall, about 1.4 seconds, as shown by the blue dots in part (a) of Fig. 2. A natural assumption is that as the sphere gets lighter, it falls more slowly, perhaps because air resistance is more important for lighter objects. An almost automatic interpolation as shown in part (b). But if we actually do the experiment,[12] we find a very different result as shown in part (c): air resistance has almost no noticeable effect until we get down to less than a few percent of the iron sphere's mass. Aluminum and wood, both much lighter than the iron, hit the ground at almost the identical instant. Neglecting air resistance (pretending you're in a vacuum) works much better for everyday falling objects than we might expect.

Another example is the Hooke's law (simple) harmonic oscillator. If you think about a real spring, like the one shown in the quiz problem in the supplementary materials, it's quite clear that Hooke's law, $T = k\Delta L$, is going to be absolutely awful if you try to compress the spring any significant fraction of its length. The coils will touch and it will be MUCH harder to reduce its total length than to stretch it. If you stretch it a lot, the coils will begin to bend instead of twist and eventually straighten the spring into a straight wire — which will behave quite differently from the simple law. Yet simple harmonic motion derived from the Hooke's law toy model is a great starting point for a wide variety of small oscillations (and works essentially perfectly for photons).

The success of surprisingly simple toy models is a bonus; a place where physics gives us back more than we expected and more than we had any right to hope for!

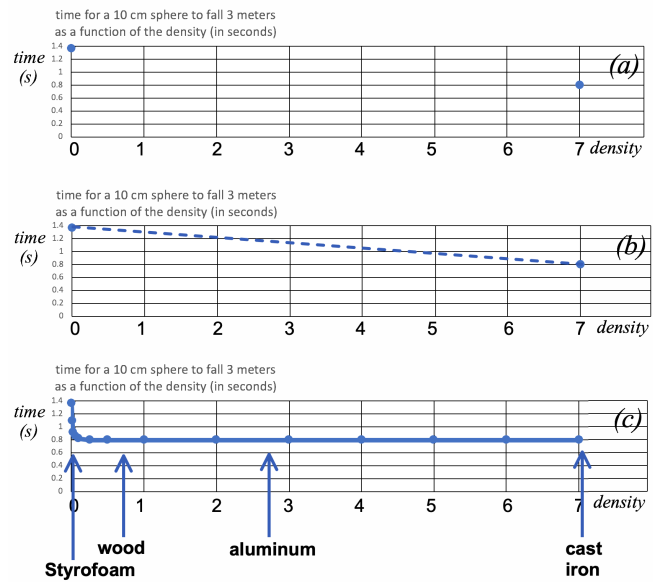

*Figure 2. The time it takes 10 cm diameter spheres of different densities to fall a distance of 3 m.*

## Toy models need to be done as part of an overall modeling strategy

Toy models need to be done not as an end in themselves but as part of a broader approach to modeling. There are two reasons for this, one positive and one negative.

- Toy models are analysis tools for approaching more complex problems.
- Toy models can inhibit making the physics/math conceptual blend.

*Toy models are analysis tools for approaching more complex problems.*

Many real-world phenomena include lots of competing effects. Making sense of them, figuring out what matters most, and how to approach them can be challenging. Toy models are not just a way of learning to build the blend; they are an *analytical tool* for approaching a complex system.

Considering what each phenomenon does by itself in a system can be a good first step in figuring out how to put them together and build a more sophisticated model. While this can be difficult in real-world situations, there are many examples that lend themselves to homework problems in an introductory class — either substantial multi-part project-style problems or shorter problems given over multiple weeks that build on and extend the result of previous weeks' problems. Some examples appropriate for a class for life science students are given in the supplementary materials.

To help students learn to value toy models, it helps to include explanations that motivate the model, problems that ask students to consider and evaluate the modeling assumptions (the three legs of Fig. 1 other than processing), and having them "go meta" by thinking and talking about models and the values explicitly.

*Toy models can inhibit making the physics/math conceptual blend.*

Toy models can be very attractive since they are typically completely solvable and completely described by a set of math tools we are comfortable with. But if the toy model is not "waaaay more realistic than we have any right to expect," it can be sufficiently unrealistic to discourage a student from seeing its relevance, undermining their sense of value of physics in their profession. This is particularly true when we are teaching non-physics STEM majors such as engineers or life science students who value authenticity and relevance. (Physicists, especially theorists like myself, are often more interested in the mathematical structure of possible theories and so may be perfectly satisfied with toy models for the control they offer.)

If toy models are used without being embedded in a more realistic modeling context, they can also inhibit the blend that both students and faculty (!) make with the real world by letting them construct an "introductory science class" epistemological frame. When someone is in this mindset, they may decide, "Since this is an intro class I don't have to consider any realistic corrections." This can result in highly inappropriate and unrealistic results and strengthen the view that physics isn't of much use in real-world situations.

## Using toy models in class

It doesn't take much to add a bit to our discussions of toy models in class to motivate them and make clear that they are an important part of our physical understanding but they are not the whole story. Being more explicit about the modeling assumptions in a standard problem can help. Here's a pretty standard example from the study of the simple harmonic oscillator, framed in an unusual way.

*Example: A block, a pellet, and a spring*

A block of mass $M$ at rest on a horizontal frictionless table is attached to a rigid support by a spring of constant $k$. A clay pellet having mass $m$ and velocity $v$ strikes the block as shown in the figure and sticks to it.

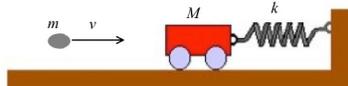

1. Determine the velocity of the block immediately after the collision.
2. Determine the amplitude of the resulting simple harmonic motion.
3. To solve this problem, you must make a number of simplifying assumptions, some of which are stated in the problem and some of which are not. Discuss the additional approximations which you had to make to solve the problem. (I can think of at least four.)

Although the question about the approximations is stated last, you really have to decide first what you are going to ignore. Unfortunately, this is often an unstated feature of many physics problems, especially ones that focus on mathematical manipulations.

The problem explicitly mentions a few simplifying assumptions:

- The table is frictionless.

This means we do not have to consider the force of friction acting on the block.

- The support (the vertical bar to which the right end of the spring is attached) is rigid.

This means that we don't have to worry about the support bending.

- The table on which the block is sliding is horizontal.

This means we don't have to worry about the cart accelerating due to gravity.

Well that's great, but there are others. Here are four simplifying assumptions that are also not only appropriate but essential to even starting a calculation:

- The spring is an ideal Hooke's law spring.
- We can ignore gravity on the pellet along its path.
- We can ignore air resistance on both pellet and cart.
- Finally, we can treat the pellet-block interaction as occurring instantaneously.

All these assumptions are typically tacit. But for a block and pellet that look like the picture and have "normal everyday" parameters — say the block is a few inches across and has a mass of a few hundred grams, the pellet is a few grams and maybe moving a few meters per second, and the spring is a typical spring that you can stretch with your two hands — these are reasonable assumptions.

As instructors, we typically make these assumptions without thinking because we have a good blend — a realistic picture of what the system looks like physically, what's important, and what's not. But students in introductory physics are rarely there yet. We need to not just be teaching toy models, but also the process of understanding what we're leaving out and how to decide when that decision is a reasonable one. Other problems asking students to consider approximations are given in the supplementary materials in EPAPS.

While we use toy models frequently in introductory physics, we rarely motivate them or discuss their limits. Here's an example of a small addition to our discussion of a toy model that can help students get better insight into what's going on.

*Example:*
*Teaching the infinite sheet of uniform charge*

While the infinite sheet of uniform charge is an immensely useful toy model, the result that the electric field points perpendicular to the plate and doesn't fall off with distance seems unnatural to many students and supports their belief that "physics isn't really about the real world."

While we can demonstrate the failure of the field to fall off with distance using an argument from dimensional analysis (the charge per unit area contains both the 1/L dimensions that the electric field needs), this neither makes the result plausible nor shows under what circumstances the model is actually relevant. If this is all the motivation we provide, it can encourage students' epistemological stance that physics is irrelevant to the real world and strengthen their belief that toy models are both ridiculous and useless.

A more detailed approach can support the idea of toy models and help build the math/physics blend by showing where the result comes from physically. We can describe how the result arises from adding up the field coming from individual bits of charge. We usually don't do this in introductory physics since it's really a vector calculus task. But it can be done conceptually to motivate the modeling and, at the same time, help students build the math/physics blend.

The key ideas are that the force between point charges adds vectorially and falls off like $1/r^2$ (key ideas coded in the Anchor Equation of Coulomb's law). Look at some point above the plane at which to calculate the E field. Each bit of charge on the plate makes a contribution proportional to $1/r^2$. As the ring gets larger, two effects come into play: (1) the angle of the E field contributions of each bit of charge gets steeper and steeper, so bits of charge on opposite sides of the ring tend to cancel more and more; (2) since the length of the ring gets longer proportional to $R$, but the $1/r^2$ distance factor to the point considered falls off like $1/(d^2 + R^2)$ the effect of each ring gets smaller as the rings get larger. The combined effect of the rings beyond about $20d$ contribution only a few percent to the total E field.

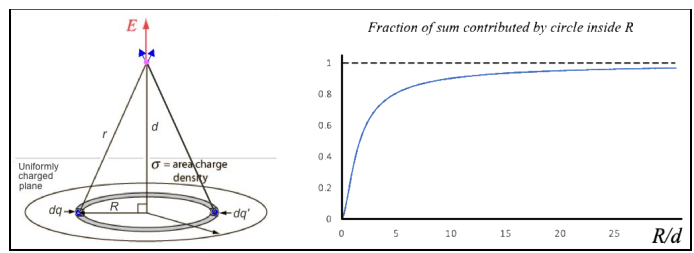

*Figure 3. Building the E field produced by a sheet of charge*

This tells us some useful things. First, the "radius of the area that contributes" grows proportional to $d$, so the amount of relevant charge grows like $d^2$. Second, the distance to the charges and the test point grows with $d$ so their effect falls off like $1/d^2$. The two effects cancel to give the constant E field independent of $d$. This provides a conceptual mechanism that makes the result plausible.

Finally, it shows when our toy model is reasonable to use. As long as the edges are further away from the point below the point we are considering by about $20d$, we can treat a finite flat sheet of charge as if it were infinite. So if we are close enough to a flat sheet, the model works.

*Example: Meta-questions*

It helps students learn to value models by having them think about them explicitly. Here's an essay question I give on an exam or as homework that illustrates the idea.

> In this class we often talk about "models". Explain what you think we mean by the term and discuss when one might be useful. Then give an example of a model from this class and one from a biology class and discuss their value. *Note: This is an essay question. Your answer will be judged not solely on its correctness, but for its depth, coherence, and clarity.*

More examples are given in the supplementary materials.

## Digging deeper: Research resources

The modern push towards including more explicit discussions of modeling was begun by David Hestenes. For some deep insights into modeling, check out his groundbreaking paper in the AJP.[13] Also see the later papers of the Modeling Group for details.[14] For an excellent overview of general modeling theory and a good introduction to models in biology, see Svoboda (Gouvea) & Passmore.[15] For more detailed discussion of the epistemological resource conflict between what introductory physics expects and what life science students are used to, see Meredith & Redish or Redish & Cooke.[16] For an excellent discussion of how engineering students' responses to toy models can lead to negative affect, see Gupta, Elby, and Danielak.[17]

## Instructional resources

The modeling group has links to a lot of resources for teaching modeling.[18] A valuable tool for modeling they introduced is the *System Schema*.[19] This focuses the first step of any problem as identifying what you are going to consider objects and what their interactions are. This leads naturally as a next step to free body diagrams in Newtonian Mechanics, and system analysis in thermodynamics. An example is given in the supplementary materials in EPAPS.

Many of the ideas for this series were developed in the context of studying physics learning in a class for life-science majors. A number of problems and activities focusing explicitly on modeling are offered in the supplementary materials to this paper. A more extensive collection of readings and activities from this project on the topic of modeling and toy models is available at the *Living Physics Portal*,[20] search "Model building."

## Acknowledgements

I would like to thank the members of the UMd PERG over the last two decades for discussion on these issues. I particularly thank Vashti Sawtelle for introducing me to the System Schema and convincing me of its value and Todd Cooke and Julia Svoboda Gouvea for conversations about modeling in biology. The work has been supported in part by a grant from the Howard Hughes Medical Institute and NSF grants 1504366 and 1624478.

---

[1] E. Redish, Using Math in Physics - Overview, preprint, and subsequent papers.

[2] American Physical Society, award for Excellence in Education (2014), American Modeling Teachers' Association;

[3] Physics majors will often cut us a lot of slack even if what we're doing doesn't make sense to them, but their lack of this resource can hinder their progress in understanding the physics they are learning.

[4] https://en.wikipedia.org/wiki/Spherical_cow

[5] *Xkcd,* by Randall Munro. https://xkcd.com/669/

[6] This is not the only way to build mathematical representations. For example, the diagram can run the other way, with mathematical structures that emerge from processing leading to new physical concepts and ways to look at the world. See paper 12 in this series (I hope): Things of physics, things of math

[7] E. Redish, Using math in physics - 1. Dimensional analysis.

[8] E. Redish, Using math in physics - 2. Estimation, preprint.

[9] E. Redish, Using math in science - 3. Anchor equations, preprint

[10] Except perhaps in an introductory class focused on computational tools!

[11] E. Redish, Using math in science - 2. Estimations, preprint

[12] This is not actual data, but a calculation done with a one-stage more sophisticated toy model: adding in an inertial drag force proportional to the square of the velocity. This agrees well with my qualitative experience in classroom demonstrations.

[13] D. Hestenes, Toward a modeling theory of physics instruction, Am. J. Phys. 55:5 (1987) 440-454.

[14] D. Hestenes, et al. (2011). A Graduate Program for High school Physics and Physical Science Teachers, *American Journal of Physics* **79**:9, p.971-979; J. Jackson, L. Dukerich, & D. Hestenes, Modeling Instruction: An Effective Model for Science Education, *Science Educator* **17:**1 (2008) 10-17.

[15] J. Svoboda & C. Passmore, The strategies of modeling in biology education, Sci. & Educ. DOI 10.1007/s11191-011-9425-5 (2011) 24 pp.

[16] D. Meredith & E. Redish, Reinventing physics for life science majors, Phys. Today 66:7 (2013) 38-43; E. Redish & T. Cooke, Learning each other's ropes: Negotiating interdisciplinary authenticity, CBE-LSE 12 (6/3/2013) 175-186.

[17] A. Gupta, A. Elby, & B. Danielak, Exploring the entanglement of personal epistemologies and emotions in students' thinking, Phys. Rev. PER 14 (2018) 010129, 22 pp.

[18] http://modeling.asu.edu/Projects-Resources.html

[19] L. Turner, System Schemas, Phys.Teach. 41:7 (2003) 404-408; B. Hinrichs, Using the System Schema Representational Tool to Promote Student Understanding of Newton's Third Law, AIP Conference Proceedings 790, 117 (2005); https://doi.org/10.1063/1.2084715.

[20] *The Living Physics Portal*, https://www.livingphysicsportal.org/